\documentclass[pre,twocolumn,showpacs]{revtex4}
\usepackage{epsfig}
\usepackage{amssymb}

\newcommand{\g}[1]{\mbox{\boldmath ${#1}$}}

\begin{document}

\title{The hodograph method applicability in the problem of long-scale 
nonlinear dynamics of a thin vortex filament near a flat boundary}
\author{V.P. Ruban}
\email{ruban@itp.ac.ru}
\affiliation{L.D.Landau Institute for Theoretical Physics,
2 Kosygin Street, 119334 Moscow, Russia}

\date{\today}

\begin{abstract}
Hamiltonian dynamics of a thin vortex filament in ideal incompressible
fluid near a flat fixed boundary is considered at the conditions that 
at any point of the curve determining shape of the filament the angle 
between tangent vector and the boundary plane is small, 
also the distance from a point on the curve to the plane is  small 
in comparison with the curvature radius. The dynamics is shown to be 
effectively described by a nonlinear system of two (1+1)-dimensional 
partial differential equations. 
The hodograph transformation reduces that system to a single linear
differential equation of the second order with  separable variables.
Simple solutions of the linear equation are investigated at real values of
spectral parameter $\lambda$ when the filament projection on the boundary
plane has shape of a two-branch spiral or a smoothed angle, depending 
on the sign of $\lambda$.
\end{abstract}


\maketitle

\section{Introduction}
It is a well known fact that solutions of equations determining the motion of
a homogeneous inviscid fluid possess a remarkable property --- the lines of
the vorticity field are frozen-in \cite{LL6,Arnold,Saffman,Chorin}. 
Mathematical reason for this is the so called relabeling symmetry 
of fluids that provides necessary conditions for the Noether theorem
applicability and results in infinite number of the conservation laws
\cite{ZK97,Salmon,PadMor,IL,R99,R2000PRD,R2001PRE} .
Due to this basic property, in the framework of ideal hydrodynamics 
such flows are possible where during sufficiently long time interval
the vorticity is concentrated in quasi-one-dimensional structures,
vortex filaments, that fill a small part of entire bulk of the fluid.
The motion of vortex filaments is very interesting problem both from 
theoretical and practical viewpoints and is a classical subject of 
hydrodynamics (see, for instance, \cite{Saffman,Chorin,R2000PRD,
RP2001PRD, R2001PRE,RPR2001PRE,Lough,HongZhou,Wang,Zakharov99} 
and references therein for various analytical and numerical approaches 
to this problem). In general case analytical study 
in this field is highly complicated because of
several reasons, the main of them being non-locality and nonlinearity of
governing equations of motion. A less significant trouble seems to be the
necessity of some regularization procedure for the Hamiltonian functional
(the total energy) of the system in the limit of ``infinitely thin'' vortex
filaments, since a logarithmic divergency takes place in some
observable physical quantities (for instance, in the velocity of
displacement of curved pieces of the filament) 
as the thickness decreases. However, in few limit cases the dynamics of a 
single vortex filament  can turn out to be effectively integrable. 
A known and very interesting example of such integrable system is a slender 
non-stretched vortex filament (in the boundless 
three-dimensional (3D) space filled by an ideal fluid) in the so called
localized induction approximation (LIA), when in the energy of the
filament only logarithmically large contributions from interaction of
adjacent pieces are taken into account. In this approximation the 
Hamiltonian is simply proportional to the length of the filament, 
resulting in conservation this quantity, thus application of the so called 
Hasimoto transformation \cite{Hasimoto,NSW1992} is appropriate and reduces
the problem to (1+1)-dimensional nonlinear Schroedinger equation that is
known to be integrable by the inverse scattering method \cite{InScTr}.

In present work another integrable case in vortex dynamics is recognized, the
long-scale motion of a thin vortex filament near a flat fixed boundary.
Mathematically the problem of a single filament in a half-space is equivalent
to the problem of two symmetric filaments in the boundless space, that allows
us simplify some further calculations. Our immediate purpose will be to
consider the configurations of the vortex filament that satisfy the following
conditions:

a) the angle is everywhere small between the tangent vector on the curve
determining the shape of the filament and the boundary plane;

b) the distance from an arbitrary point of the curve to the plane is small
comparatively to the curvature radius at the given point but large in
comparison with the thickness of the filament;

c) the filament projection on the boundary plane does not have
self-intersections or closely approaching one-to-another different pieces.

In these conditions the system dynamics is known to be unstable (the so
called Crow instability \cite{Crow}), with the instability increment 
directly proportional to the wave number of (some small) long-scale
perturbation of the filament shape. It is a well known fact that such 
dependence of the increment is usual for a class of local $(2\times 2)$ 
partial differential systems that can be exactly linearized by 
so called hodograph transformation \cite{LL6} exchanging 
dependent and independent variables.
This observation has served as a weighty reason to look for a natural
local nonlinear approximation in description of the long-scale dynamics of a
vortex filament near a flat boundary and to examine that approximation for the
hodograph method applicability. As the result, a consistent derivation of the
corresponding local approximate equations of motion has been performed, and also
the fact has been demonstrated that the nonlinear partial differential system 
for two functions $\rho$ and $\vartheta$ determining the shape of the filament 
and depending on the time moment $t$ and on the Cartesian coordinate $x$ 
is reduced by hodograph transformation to a linear equation.
Moreover, it is possible to choose the pair of new independent variables in 
such a manner that in the linear partial differential equation for the 
function $t(\rho,\vartheta)$ the coefficients will not depend on  
$\vartheta$-variable. For this purpose it is convenient to define 
$\rho$-variable as double distance from a filament point to the 
boundary plane $y=0$, while $\vartheta$-variable will be the angle between 
$x$-direction and the tangent to the filament projection on 
$(x,z)$-plane. Obviously, an explicit dependence of the coefficients on 
$\vartheta$  will be absent due to the symmetry of the system with respect to
rotations in $(x,z)$-plane. Therefore separation of the variables will be
possible and most simple solutions will have the form
\begin{equation}\label{t_lambda}
t_\lambda(\rho,\vartheta)=\mbox{Re}\left\{
T_\lambda(\rho)\Theta_\lambda(\vartheta)\right\},
\end{equation}
where $\lambda$ is an arbitrary complex parameter, and the complex function 
$\Theta_\lambda(\vartheta)$ satisfies the simple equation 
\begin{equation}\label{equation_for_Theta}
\Theta_\lambda''(\vartheta)=\lambda\Theta_\lambda(\vartheta).
\end{equation}
To find the complex function $T_\lambda(\rho)$, it will be necessary to solve
some ordinary linear differential equation of the second order with variable
coefficients that will be considered in appropriate section of this paper. 
The corresponding geometrical configurations of the vortex filament strongly 
depend on $\lambda$.
In particular, it will be shown the solutions (\ref{t_lambda}) with 
real $\lambda<-1$ describe such a shape of the (moving) vortex filament
that its projection on $(x,z)$-plane has two asymptotes with the angle
between them $\Delta\vartheta=\pi(1-1/\sqrt{-\lambda})$, while in the case 
$\lambda>0$ the projection has the shape of a two-branch
spiral (see the figures).

This article is organized as follows. In section II a necessary brief review 
is given concerning the Hamiltonian formalism adopted 
to the problem of frozen-in vorticity, since this approach is the most clear and 
compact way to treat ideal flows. Then approximate local equations of motion 
for a vortex filament near a flat boundary are derived.
In section III we demonstrate applicability of the hodograph method and 
introduce variables that are most convenient for the particular problem. 
Section IV is devoted to investigation of simple solutions obtained by 
the variables separation in the governing linear equation.

\section{Long-scale local approximation}
% in dynamics of a vortex filament}

Existence itself of the ideal-hydrodynamic solutions in the form of
quasi-one-dimensional vortex structures (vortex filaments) filling just a
small part of the total fluid bulk is closely connected with the 
freezing-in property of the vortex lines \cite{LL6,Arnold,Saffman,Chorin,
ZK97,R2001PRE,RPR2001PRE,Salmon,PadMor,IL,R99,R2000PRD,RP2001PRD}. 
Mathematically this property is expressed by the special form of the equation 
of motion for the divergence-free vorticity field 
$\g{\Omega}(\g{r},t)=\mbox{curl\,}\g{v}(\g{r},t)$,
\begin{equation}\label{frozen-in_Omega}
\g{\Omega}_t=\mbox{curl\,}[\g{v}\times\g{\Omega}],
\end{equation}
where $\g{v}(\g{r},t)$ is the velocity field. Since in this article
we consider incompressible flows, we may write
\begin{equation}\label{v_Omega}
\g{v}=\mbox{curl\,}^{-1}\g{\Omega}=\mbox{curl\,}(-\Delta)^{-1}\g{\Omega},
\end{equation}
where $\Delta$ is the 3D Laplace operator. As it is known, the action of the
inverse nonlocal operator $\Delta^{-1}$ on an arbitrary function $f(\g{r})$
is given by the formula
\begin{equation}
-\Delta^{-1}f(\g{r})=\int G(|\g{r}-\g{r}_1|)f(\g{r}_1)d\g{r}_1,
\end{equation}
where
$$
G(r)=\frac{1}{4\pi r}
$$ 
is the Green function of the $(-\Delta)$ operator in the boundless space.
Hamiltonian noncanonical structure \cite{ZK97} of the equations of
ideal incompressible hydrodynamics is based on the relation
\begin{equation}\label{v_Omega_Hamiltonian}
\g{v}=\mbox{curl\,}\left(\frac{\delta{\cal H}}{\delta\g{\Omega}}\right),
\end{equation}
where the Hamiltonian functional ${\cal H}\{\g{\Omega}\}$ is the kinetic 
energy of a homogeneous incompressible fluid (with unit density) expressed 
through the vorticity,
\begin{equation}\label{H_Euler}
{\cal H}\{\g{\Omega}\}=
\frac{1}{2}\int \g{\Omega}\cdot(-\Delta)^{-1}\g{\Omega}\,d\g{r}.
\end{equation} 

Our approach to investigation of the vortex filament motion is based on 
representation of the ideal homogeneous fluid flows in terms of the frozen-in 
vortex lines, as described, for instance, in 
\cite{R2000PRD,R2001PRE, RP2001PRD, RPR2001PRE}. 
The special form (\ref{frozen-in_Omega}) of the equation of motion allows 
us  express the vorticity field $\g{\Omega}(\g{r},t)$ in a self-consistent
manner through the shape of the vortex lines (the so called formalism of 
vortex lines),
\begin{equation}
\g{\Omega}(\g{r},t)=\int_{\cal N} d^2\nu\oint 
\delta(\g{r}-\g{R}(\nu,\xi,t))\g{R}_{\xi}(\nu,\xi,t)d\xi,
\end{equation}
where $\delta(\dots)$ is the 3D delta-function, ${\cal N}$  is some 2D 
manifold of labels enumerating the vortex lines (${\cal N}$ is determined by 
topological  properties of a particular flow), $\nu\in{\cal N}$ is a label 
of an individual vortex line, $\xi$ is an arbitrary longitudinal parameter 
along the line. 
What is important, the dynamics of the line shape 
$\g{R}(\nu,\xi,t)=(X(\nu,\xi,t),Y(\nu,\xi,t),Z(\nu,\xi,t))$ is determined by
the variational principle
$$
\delta\left[\int{\cal L}dt\right]/\delta\g{R}(\nu,\xi,t)=0,
$$ 
with the Lagrangian of the form
\begin{equation}
{\cal L}=\int_{\cal N} d^2\nu\oint
([\g{R}_{\xi}\times\g{R}_t]\cdot\g{D}(\g{R}))d\xi \,\,
-{\cal H}\{\g{\Omega}\{\g{R}\}\},
\end{equation}
where the vector function $\g{D}(\g{R})$ in the case of incompressible flows
must satisfy the condition
\begin{equation}
(\nabla_{\mbox{\scriptsize \g{R}}}\cdot\g{D}(\g{R}))=1.
\end{equation}
Below we choose $\g{D}(\g{R})=(0,Y,0)$.

Since we are going to deal with a very thin vortex filament, we will neglect 
the $\nu$-dependence of the shapes of individual vortex lines constituting 
the filament. By doing this step we exclude from further consideration all 
effects related to finite variable cross-section and longitudinal flows 
inside the filament. Thus, we consider an ``infinitely narrow'' vortex string 
with a shape $\g{R}(\xi,t)$ and with a finite circulation 
$\Gamma=\int_{\cal N} d^2\nu$.
However, the Hamiltonian of such singular filament diverges logarithmically,
\begin{equation}\label{Hamiltonian_singular}
{\cal H}_{\Gamma}\{\g{R}(\xi)\}=\frac{\Gamma^2}{8\pi}\oint\!\!\oint
\frac {\g{R}'(\xi_1)\cdot\g{R}'(\xi_2)\,\,d\xi_1\,d\xi_2}
{|\g{R}(\xi_1)-\g{R}(\xi_2)|}
\,\,\to \infty
\end{equation}
In order to regularize this double integral, it is possible, as a variant, to
modify the Green function \cite{RPR2001PRE}. For example, 
instead of the singular function $G\propto 1/r$ one can use a smooth function 
like $G_a\propto 1/\sqrt{r^2+a^2}$ or some other appropriate expression 
depending on a parameter $a$.
It should be emphasized that relation $\g{\Omega}=\mbox{curl }\g{v}$ is
exactly satisfied only in the original non-regularized system, 
but in the case of a finite $a$ it is not valid on distances of order 
$a$ from the singular vortex string. Thus, the meaning of vorticity
in regularized models is not so simple, but nevertheless, relation 
(\ref{v_Omega_Hamiltonian}) remains valid in any case.
Relatively small parameter $a$ in regularized models serves to imitate 
a finite width of vortex filament in the usual (non-regularized)
hydrodynamics. The energy of the string turns out to be logarithmically large,
\begin{equation}
{\cal H}_{\Gamma}\{\g{R}(\xi)\}\approx
\frac{\Gamma^2}{4\pi}\oint |\g{R}'(\xi)|
\log\left(\frac{\Lambda(\g{R}(\xi))}{a}\right)d\xi,
\end{equation}
where $\Lambda(\g{R})$ is a typical scale depending on a given problem
(in particular, the usual LIA uses $\Lambda=$ const $\gg a$).
In our case we consider two symmetric vortex filaments in the long-scale 
limit, when direction of the tangent vector varies weakly on
a length of order $Y$. For such configurations, the energy concentrated in
the half-space $y>0$ is approximately equal to the following expression
\begin{equation}\label{LIA_generalized}
{\cal H}_{\Gamma}\approx \frac{\Gamma^2}{4\pi}\oint\sqrt{X'^2+Y'^2+Z'^2}\,
\log(2Y/a) d\xi.
\end{equation}
This local Hamiltonian is able to provide qualitatively correct  
dynamics of the filament down to longitudinal scales of order $Y$ where 
perturbations become stable and where non-locality comes to play. 
Unfortunately, we do not have a simple method 
to  treat the Hamiltonian (\ref{LIA_generalized}) analytically, that is why 
we will consider only very large scales ($L \gg Y$) and thus suppose the 
slope of the tangent vector to the boundary 
plane to be negligibly small (this means  $Y'^2\ll X'^2+Z'^2$). 
Then, choosing as a longitudinal parameter $\xi$ simply the Cartesian 
coordinate $x$, we have the following approximate Lagrangian:
\begin{equation}\label{approx_Lagrangian}
{\cal L}\approx \Gamma \int \left\{-Y \dot Z 
-\frac{\Gamma}{4\pi}\sqrt{1+Z'^2}
\log\left({2Y}/{a}\right)\right\}dx,
\end{equation}
where the functions $Y$ and $Z$ depend on $x$ and $t$, 
$\dot Z\equiv\partial_t Z$, $Z'\equiv\partial_x Z$.
Having neglected the term $Y'^2$ under the square root, we sacrifice correct
behaviour of perturbations with wave-lengths of order $Y$, but instead
we obtain exactly solvable system, as it will be shown below.

Let us say a few words about geometrical meaning of the second term in
r.h.s. of the expression (\ref{approx_Lagrangian}). 
Since we study the very long-scale limit, 
locally the flow under  consideration looks almost like a two-dimensional 
flow with a small vortex at the distance $Y$ from the straight boundary, 
and the expression $({\Gamma}/{4\pi})\log(2Y/a)$ is just the energy of such 
2D flow per unit length in the third (longitudinal) direction, 
while the multiplier $\sqrt{1+Z'^2}\,dx$ gives the arc-length element 
in the longitudinal direction.

Now for simplicity we take new time and length scales to satisfy $a=1$ and 
${\Gamma}/{2\pi}=1$.
After that we introduce new quantities $\rho(x,t)=2Y(x,t)$ and 
$\mu(x,t)=\partial_x Z(x,t)$, and also the function $H(\rho,\mu)$, 
\begin{equation}\label{H_rm}
H(\rho,\mu)=F(\rho)\,\sqrt{1+\mu^2},
\end{equation}
where
\begin{equation}\label{F_log}
F(\rho)=\log \rho.
\end{equation}
The corresponding equations of motion then can be written in the following
remarkable general form:
\begin{eqnarray}
\mu={\partial_ x}Z,&&\label{mu_definition}\\
{\partial_t}\rho+{\partial_x}H_\mu(\rho,\mu)=0,&&\label{contin_general}\\
{\partial_t}Z+H_\rho(\rho,\mu)=0.&&\label{Bernoulli_general}
\end{eqnarray}
More explicitly, the last two equations are
\begin{eqnarray}
\rho_t+\frac{\partial}{\partial x}\Bigg [\frac{F(\rho) Z_x}{\sqrt{1+ Z_x^2}}\Bigg ]=0,&&
\label{rho_t}\\
{\partial_t}Z+F'(\rho)\sqrt{1+Z_x^2}=0.&&\label{Z_t}
\end{eqnarray}
These equations have a simple geometrical treatment. Indeed, Eq.(\ref{Z_t}) 
means if we consider the dynamics of the filament projection on $(x,z)$-plane, 
then we see an element of the projection moving in
the normal to the projection tangent direction with the velocity depending only
on $\rho$ and equal to $F'(\rho)$. Simultaneously, in $y$-direction the element
of the filament moves with the velocity proportional to the $(x,z)$-projection
curvature multiplied by the function $F(\rho)$, 
as Eq.(\ref{rho_t}) shows.

It is interesting to note that an analogous consideration can give us also
long-scale Hamiltonian equations of motion for a thin vortex filament in a 
slab of an ideal fluid between two parallel fixed boundaries at $y=-d/2$ and 
$y=+d/2$. One has just to define the $\rho$-variable by the formula 
$\rho=(\pi/d)y$ and make in Eq.(\ref{H_rm}) substitution 
$F(\rho)\mapsto F^{(\epsilon)}(\rho)$ where
\begin{equation}\label{F_epsilon}
F^{(\epsilon)}(\rho)=\log\left(\frac{\cos\rho}{\epsilon}\right),
\end{equation}
with a small dimensionless parameter $\epsilon$.

\section{Hodograph method}

It is known that any nonlinear system of the form 
(\ref{mu_definition})-(\ref{Bernoulli_general}) can be locally reduced to a
linear equation if we take $\rho$ and $\mu$ as new independent variables (this
is the so called hodograph method; see, for instance, \cite{LL6} where a
particular case is discussed, the 1D gas dynamics, with 
$H(\rho,\mu)=\rho\mu^2/2+\varepsilon(\rho)$, where $\rho$, $\mu$, 
$\varepsilon(\rho)$ are the gas density, gas velocity, and the internal energy
density respectively). Indeed, as from equations (\ref{mu_definition}) and 
(\ref{Bernoulli_general}) we see the relation
$$
dZ=\mu\, dx -H_\rho \,dt,
$$
it is convenient to introduce the auxiliary function $\chi(\rho,\mu)$,
\begin{equation}\label{chi_definition}
\chi=Z- x\mu + tH_\rho
\end{equation}
in order to obtain
$$
d\chi=-x\,d\mu +tH_{\rho\rho}\,d\rho +t H_{\rho\mu}\,d\mu.
$$
From the above expression we easily derive
\begin{equation}\label{TX}
t=\frac{\chi_\rho}{H_{\rho\rho}}, \qquad
x=H_{\rho\mu}\frac{\chi_\rho}{H_{\rho\rho}}-\chi_\mu\label{X}.
\end{equation}
After that we rewrite Eq.(\ref{contin_general}) in the form
$$
\frac{\partial(\rho,x)}{\partial(t,x)}-
H_{\mu\rho}\frac{\partial(\rho,t)}{\partial(t,x)}-
H_{\mu\mu}\frac{\partial(\mu,t)}{\partial(t,x)}=0
$$
and multiply it by the Jacobian ${\partial(t,x)}/{\partial(\rho,\mu)}$,
$$
\frac{\partial(\rho,x)}{\partial(\rho,\mu)}-
H_{\mu\rho}\frac{\partial(\rho,t)}{\partial(\rho,\mu)}-
H_{\mu\mu}\frac{\partial(\mu,t)}{\partial(\rho,\mu)}=0.
$$
Thus, now we have
\begin{equation}\label{x_mu}
x_\mu=H_{\mu\rho}t_\mu-H_{\mu\mu}t_\rho.
\end{equation}
Substitution of the relations (\ref{TX}) into this equation and subsequent
simplification give us the linear partial differential equation of the second
order for the function $\chi(\rho,\mu)$,
\begin{equation}\label{general_linear_equation_for_chi}
({H_{\mu\mu}}\chi_{\rho}/{H_{\rho\rho}})_\rho 
-\chi_{\mu\mu}=0.
\end{equation}

As the function  $H(\rho,\mu)$ has the special form (\ref{H_rm}), it is
convenient to change variables,
\begin{equation}\label{mu_and_u}
\mu=\tan\vartheta,\qquad 
\chi(\rho,\vartheta)=-\frac{u(\rho,\vartheta)}{\cos\vartheta},
\end{equation}
where $\vartheta$ is the angle in $(x,z)$-plane  between $x$-direction and the
tangent to the corresponding projection of the vortex filament. As the result,
the relations  (\ref{chi_definition}) and (\ref{TX}) will be rewritten in the
form
\begin{eqnarray}\label{time_u}
&&t=-\frac{u_\rho}{F''(\rho)} \\
\label{x_u}
&&x=u_\vartheta\cos\vartheta+
\left(u-\frac{F'(\rho)}{F''(\rho)}u_\rho\right)\sin\vartheta,\\
\label{Z_u}
&&Z=u_\vartheta\sin\vartheta-
\left(u-\frac{F'(\rho)}{F''(\rho)}u_\rho\right)\cos\vartheta,
\end{eqnarray}
and the coefficients of the linear equation for the function 
$u(\rho,\vartheta)$ will not depend on $\vartheta$-variable,
\begin{equation}\label{equation_for_u_rho_theta}
\frac{\partial}{\partial\rho}\left(\frac{F(\rho)}{F''(\rho)}u_{\rho}\right)
-(u_{\vartheta\vartheta}+u)=0.
\end{equation}
The same is true for the coefficients of the equation determining the function 
$t(\rho,\vartheta)=-u_\rho(\rho,\vartheta)/F''(\rho)$,
\begin{equation}\label{equation_for_t_rho_theta}
F(\rho) t_{\rho\rho} +2F'(\rho) t_{\rho}-F''(\rho) t_{\vartheta\vartheta}=0.
\end{equation}

Once some particular solution of Eq.(\ref{equation_for_u_rho_theta}) is known, 
then further procedure consists in the following two steps:

i) find in terms of some parameter $\xi$
the curves of constant values of the function 
$t=-u_\rho(\rho,\vartheta)/F''(\rho)$. 
It is this point where nonlinearity comes to play, 
since we need to solve nonlinear equation; 

ii) substitute  the obtained expressions $\rho=\rho(\xi,t)$ and 
$\vartheta=\vartheta(\xi,t)$ into Eqs.(\ref{x_u}-\ref{Z_u})
and get complete description of the filament motion, 
$X=X(\xi,t)$, $Z=Z(\xi,t)$, $Y=(1/2)\rho(\xi,t)$.

Thus,  the long-scale local approximation (\ref{approx_Lagrangian}) 
turns out to be integrable in the sense it is reduced to the {\em linear} 
equation (\ref{equation_for_u_rho_theta}). However, the function 
$u(\rho,\vartheta)$ is multi-valued  in the general case. 
Therefore statement of the Cauchy problem
becomes much more complicated. Besides, the functions $F(\rho)$ and 
$F^{(\epsilon)}(\rho)$ determined by expressions (\ref{F_log}) and 
(\ref{F_epsilon}) result in {\em elliptic} linear equations as against the
usual 1D gas dynamics where the corresponding equations were {\em hyperbolic}.
Generally speaking, the ellipticity makes the Cauchy problem ill-posed in the
mathematical sense if initial data are not very smooth. However, in this
article we will not discuss these questions, instead in the following section 
we will present simple particular solutions that during some time interval 
satisfy the applicability conditions for the long-scale approximation.

%\begin{widetext}
\section{Particular solutions}
\subsection{Separation of the variables}

We are going to consider the simplest particular solutions of  
Eq.(\ref{equation_for_u_rho_theta}) obtainable by separation of the variables
\begin{equation}\label{u_lambda}
u_\lambda(\rho,\vartheta)=
\mbox{Re}\left\{U_\lambda(\rho)\Theta_\lambda(\vartheta)\right\},
\end{equation}
where $\lambda$ is an arbitrary complex spectral parameter,
\begin{equation}
\lambda=(\varkappa+ik)^2, \qquad k\ge 0,
\end{equation}
and the function $\Theta_\lambda(\vartheta)$ contains two arbitrary complex
constants $C_\lambda^+$ and $C_\lambda^-$,
\begin{equation}\label{Theta_separat}
\Theta_\lambda(\vartheta)=
C_\lambda^{+}\exp[(\varkappa+ik)\vartheta]+
C_\lambda^{-}\exp[-(\varkappa+ik)\vartheta].
\end{equation}
The motion of the vortex filament will be described by the formulas
\begin{eqnarray}\label{t_lambda_U}
t_\lambda&=&-\mbox{Re}\left\{\frac{U_\lambda'(\rho)}{F''(\rho)}
\Theta_\lambda(\vartheta)\right\}, \\
\label{x_lambda_U}
x_\lambda&=&\mbox{Re}\Big\{
U_\lambda(\rho)\Theta_\lambda'(\vartheta)\cos\vartheta \nonumber \\
&&+\Big(U_\lambda(\rho)-\frac{F'(\rho)}{F''(\rho)}U_\lambda'(\rho)\Big)
\Theta_\lambda(\vartheta)\sin\vartheta \Big\},\\
\label{Z_lambda_U}
Z_\lambda&=&\mbox{Re}\Big\{
U_\lambda(\rho)\Theta_\lambda'(\vartheta)\sin\vartheta \nonumber \\
&&-\Big(U_\lambda(\rho)-\frac{F'(\rho)}{F''(\rho)}U_\lambda'(\rho)\Big)
\Theta_\lambda(\vartheta)\cos\vartheta \Big\} .
\end{eqnarray}
The function $U_\lambda(\rho)$ must satisfy the ordinary differential equation
of the second order
\begin{equation}\label{U_lambda}
\frac{d}{d\rho}\left(\frac{F(\rho)}{F''(\rho)}U_\lambda'(\rho)\right)
-(\lambda+1)U_\lambda(\rho)=0.
\end{equation}

Let us turn a bit of attention to the special value $\lambda=-1$ of the
spectral parameter, when the solution of Eq.(\ref{U_lambda}) can be
explicitly written for any function $F(\rho)$,
\begin{equation}\label{U1}
U_{-1}(\rho)=A_{-1}\int^\rho\frac{F''(\rho_1)d\rho_1}{F(\rho_1)} +B_{-1},
\end{equation}
where $A_{-1}$ and $B_{-1}$ are arbitrary complex constants.

At $\lambda\not= -1$ it can be convenient to deal with the function
\begin{equation}
T_\lambda(\rho)=-\frac{U_\lambda'(\rho)}{F''(\rho)},
\end{equation}
that satisfies the equation
\begin{equation}\label{T_lambda}
F(\rho) T_\lambda''(\rho) +2F'(\rho) T_\lambda'(\rho)-
\lambda F''(\rho) T_\lambda(\rho)=0.
\end{equation}
In particular, Eq.(\ref{T_lambda}) is simply solvable at $\lambda=0$ (this 
solution describes the motion of a perfect vortex ring),
\begin{equation}\label{lambda0}
T_0(\rho)=A_0\int^\rho\frac{d\rho_1}{F^2(\rho_1)} +B_0.
\end{equation}

Simple manipulations with formulas (\ref{t_lambda_U}-\ref{U_lambda}) allow us 
rewrite the solutions in the form
\begin{eqnarray}
\label{t_lambda_T}
&&t_\lambda=\mbox{Re}
\Big\{T_\lambda(\rho)\Theta_\lambda(\vartheta)\Big\},\\
\label{x_lambda_T}
&&x_\lambda=\mbox{Re}\Big\{(\lambda+1)^{-1}\{
-[F(\rho)T_\lambda(\rho)]'\Theta_\lambda'(\vartheta)\cos\vartheta
\nonumber\\
&&\quad+[\lambda F'(\rho)T_\lambda(\rho)-F(\rho)T_\lambda'(\rho)]
\Theta_\lambda(\vartheta)\sin\vartheta \}\Big\},\\
\label{Z_lambda_T}
&&Z_\lambda=\mbox{Re}\Big\{(\lambda+1)^{-1}\{
-[F(\rho)T_\lambda(\rho)]'\Theta_\lambda'(\vartheta)\sin\vartheta
\nonumber\\
&&\quad-[\lambda F'(\rho)T_\lambda(\rho)-F(\rho)T_\lambda'(\rho)]
\Theta_\lambda(\vartheta)\cos\vartheta \}\Big\}.
\end{eqnarray}

\subsection{Real $\lambda$}

Let us first consider real values of the spectral parameter, 
$\lambda\in{\cal R}$, and the corresponding real functions 
$\Theta_\lambda(\vartheta)$ and $U_\lambda(\rho)$. Since 
$F(\rho) >0$, $F''(\rho) <0$, we may expect the solutions 
$U_\lambda(\rho)$ with $\lambda\gg 1$ to have a number of oscillations, the
more, the more $\lambda$ is. In the opposite case, when $\lambda < -1$,
the solutions will be a linear combination of two functions, one of
them being increasing, and other decreasing. It is sufficient to know these
general properties to get an impression concerning geometrical
configurations of the vortex filament described by 
the formulas (\ref{t_lambda_T}-\ref{Z_lambda_T}). Let us take
$\lambda=-k^2$ with $k>1$ and suppose the explicit dependence $T_{-k^2}(\rho)$
to be known and increasing at large $\rho$. For simplicity we take
$\Theta_{-k^2}(\vartheta)=\cos(k\vartheta)$ and after that resolve the relation
(\ref{t_lambda_T}) with respect to $\vartheta$,
\begin{equation}\label{theta_rt}
\vartheta=\pm\frac{1}{k}
\arccos\left[\frac{t}{T_{-k^2}(\rho)}\right].
\end{equation}
Substitution of this expression into formulas (\ref{x_lambda_T}-\ref{Z_lambda_T}) 
gives us final form of the solutions as dependences $X_{-k^2}(\rho,t)$ 
and $Z_{-k^2}(\rho,t)$,
\begin{widetext}
\begin{eqnarray}\label{x_rt}
X_{-k^2}(\rho,t)&=&\pm\frac{k^2F'(\rho)T_{-k^2}(\rho)+F(\rho)T_{-k^2}'(\rho)}{k^2-1}
\left[\frac{t}{T_{-k^2}(\rho)}\right]\,
\sin\left(\frac{1}{k}\arccos\left[\frac{t}{T_{-k^2}(\rho)}\right]\right)\nonumber \\
&&\mp\frac{k[F(\rho)T_{-k^2}(\rho)]'}{k^2-1}
\left[1-\frac{t^2}{T_{-k^2}^2(\rho)}\right]^{1/2}
\cos\left(\frac{1}{k}\arccos\left[\frac{t}{T_{-k^2}(\rho)}\right]\right),
\end{eqnarray}
\begin{eqnarray}\label{Z_rt}
Z_{-k^2}(\rho,t)&=&\mp\frac{k^2F'(\rho)T_{-k^2}(\rho)+F(\rho)T_{-k^2}'(\rho)}{k^2-1}
\left[\frac{t}{T_{-k^2}(\rho)}\right]\,
\cos\left(\frac{1}{k}\arccos\left[\frac{t}{T_{-k^2}(\rho)}\right]\right)\nonumber \\
&&\mp\frac{k[F(\rho)T_{-k^2}(\rho)]'}{k^2-1}
\left[1-\frac{t^2}{T_{-k^2}^2(\rho)}\right]^{1/2}
\sin\left(\frac{1}{k}\arccos\left[\frac{t}{T_{-k^2}(\rho)}\right]\right).
\end{eqnarray}
The $\rho$-variable in the above expressions varies in the limits from
$\rho_{min}(t)$ such that $t=T_{-k^2}(\rho_{min})$, to $+\infty$. The
corresponding curve in $(x,z)$-plane is a smoothed angle 
$\Delta\vartheta=\pi(1-1/k)$ [see Figs.(\ref{k_sqrt9}-\ref{k_sqrt10}) 
where for the case $F(\rho)=\log \rho$ the filament shape is shown 
at several time moments, $t_{n+1}-t_n=$ const].
\begin{figure}
\begin{center}
  \epsfig{file=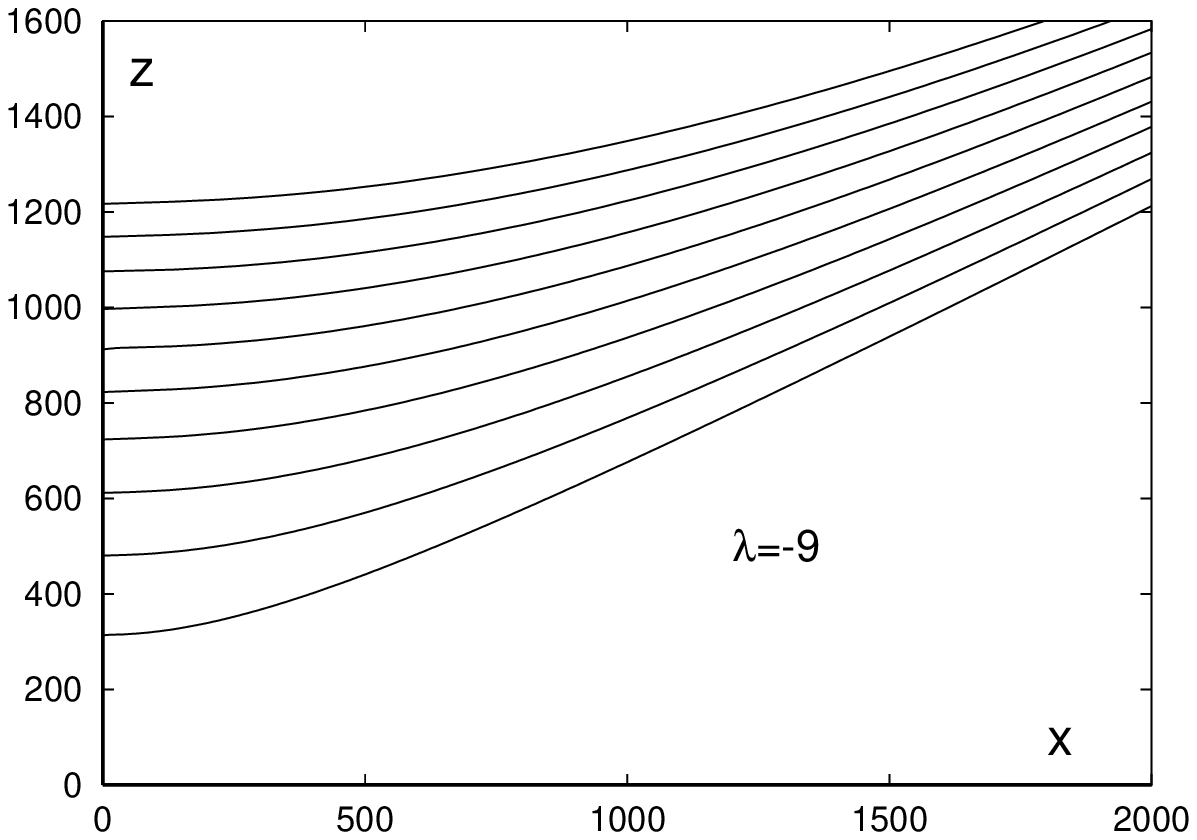,width=85mm}
  \epsfig{file=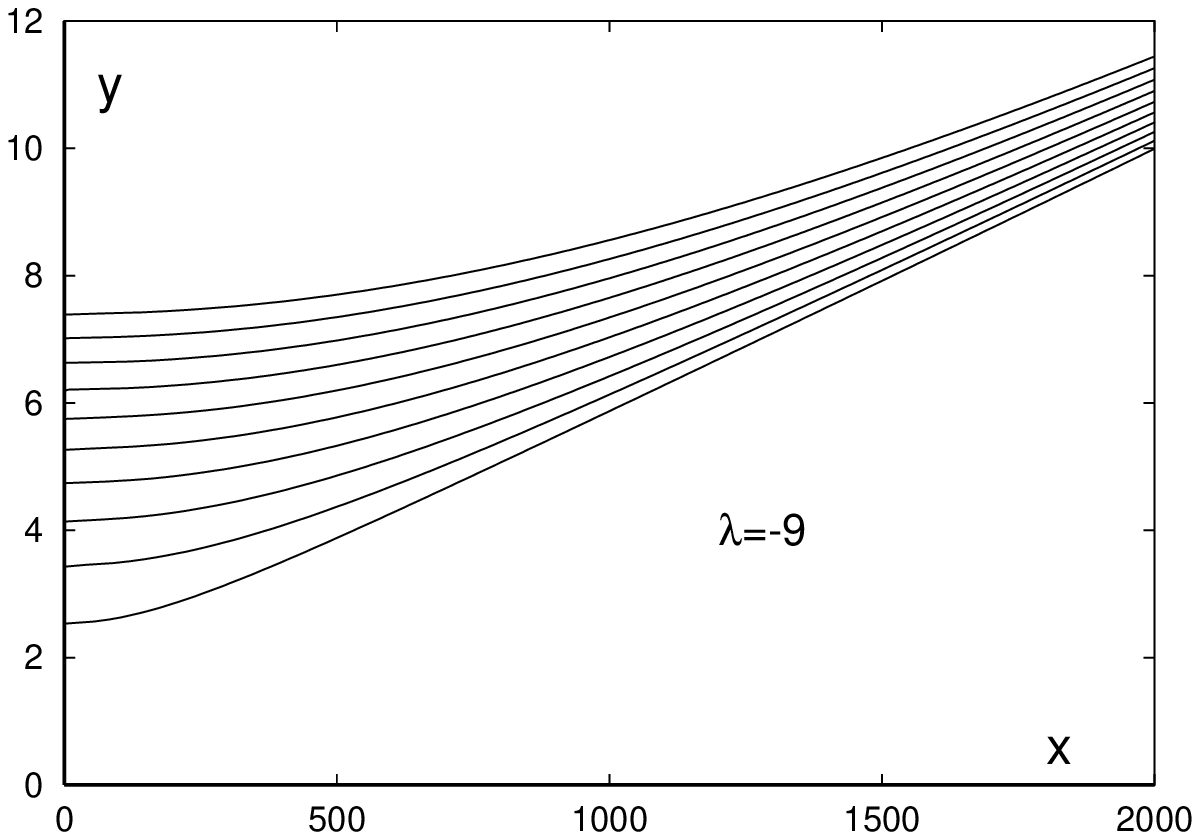,width=85mm}  
\end{center}
\caption{\small Solution for $\lambda=-9$.} 
\label{k_sqrt9}
\end{figure}
\begin{figure}
\begin{center}
  \epsfig{file=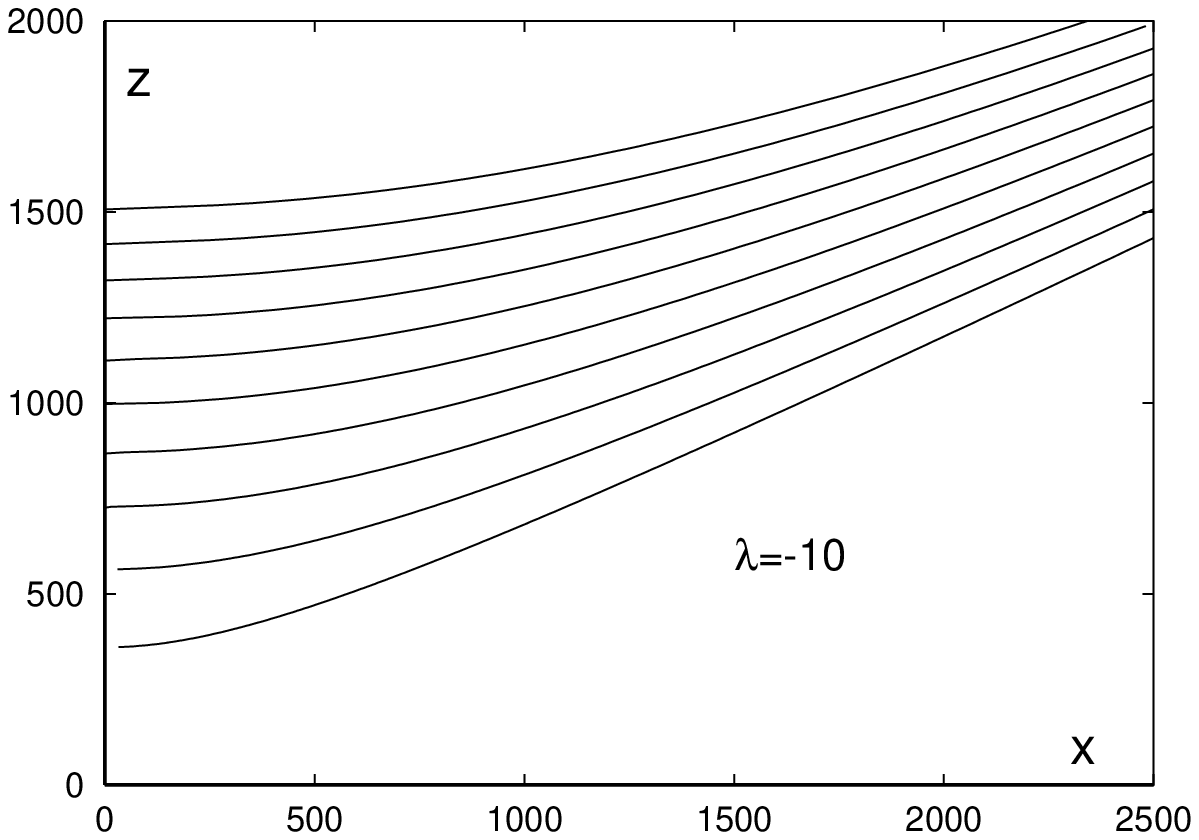,width=85mm}
  \epsfig{file=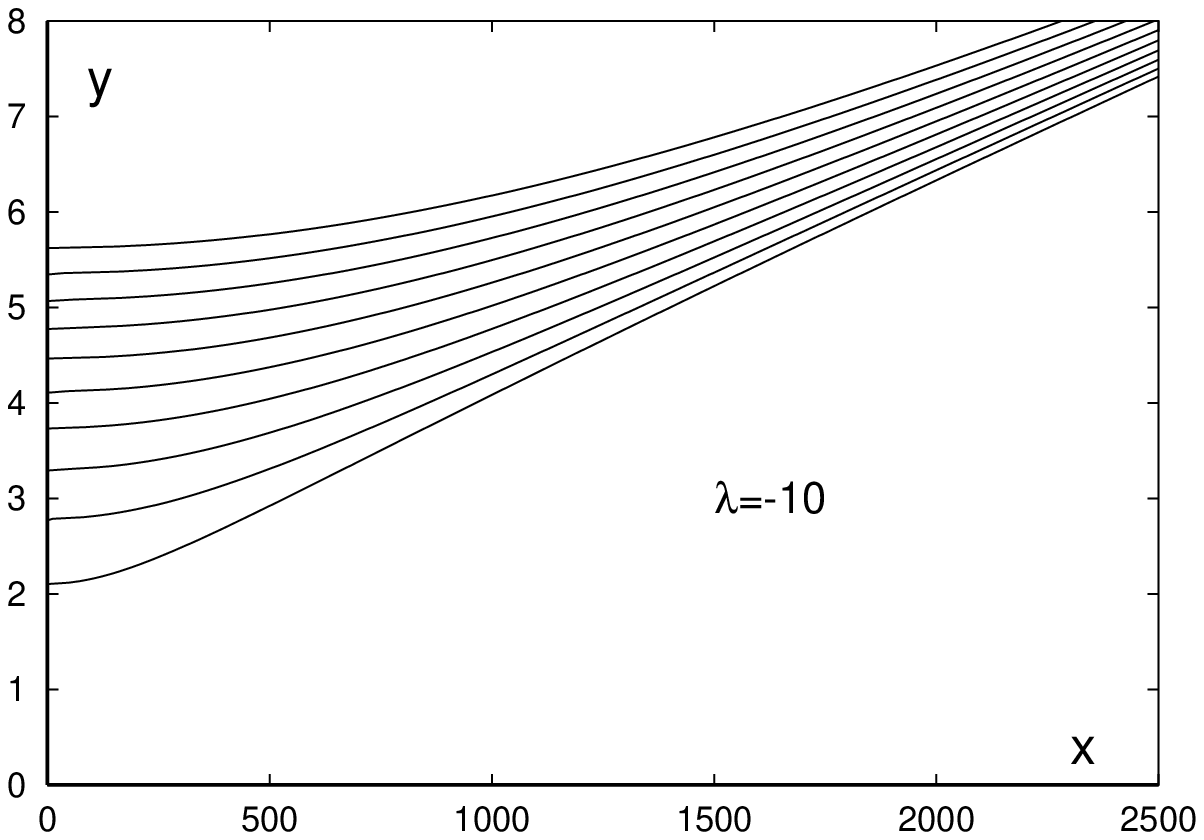,width=85mm}  
\end{center}
\caption{\small Solution for $\lambda=-10$.}
\label{k_sqrt10}
\end{figure}
Completely different form is obtained at $\lambda=\varkappa^2$, two-branch
spirals [see Fig.(\ref{spirals})]. 
Let us take $\Theta_{\varkappa^2}(\vartheta)=\exp(\varkappa \vartheta)$.
Then
\begin{equation}\label{theta_spiral}
\vartheta=\frac{1}{\varkappa}
\log\left[\frac{t}{T_{\varkappa^2}(\rho)}\right],
\end{equation}
\begin{eqnarray}\label{x_spiral}
X_{\varkappa^2}(\rho,t)&=&
\frac{t}{T_{\varkappa^2}(\rho)}\Biggr\{
\frac{\varkappa^2 F'(\rho)T_{\varkappa^2}(\rho)
-F(\rho)T_{\varkappa^2}'(\rho)}{\varkappa^2+1}
\sin\left(\frac{1}{\varkappa}
\log\left[\frac{t}{T_{\varkappa^2}(\rho)}\right]\right)\nonumber \\
&&\qquad\qquad\qquad\qquad
-\frac{\varkappa[F(\rho)T_{\varkappa^2}(\rho)]'}{\varkappa^2+1}
\cos\left(\frac{1}{\varkappa}
\log\left[\frac{t}{T_{\varkappa^2}(\rho)}\right]\right)\Biggr\},
\end{eqnarray}
\begin{eqnarray}\label{Z_spiral}
Z_{\varkappa^2}(\rho,t)&=&
\frac{t}{T_{\varkappa^2}(\rho)}\Biggr\{
-\frac{\varkappa^2 F'(\rho)T_{\varkappa^2}(\rho)
-F(\rho)T_{\varkappa^2}'(\rho)}{\varkappa^2+1}
\cos\left(\frac{1}{\varkappa}
\log\left[\frac{t}{T_{\varkappa^2}(\rho)}\right]\right)\nonumber \\
&&\qquad\qquad\qquad\qquad
-\frac{\varkappa[F(\rho)T_{\varkappa^2}(\rho)]'}{\varkappa^2+1}
\sin\left(\frac{1}{\varkappa}
\log\left[\frac{t}{T_{\varkappa^2}(\rho)}\right]\right)\Biggr\}.
\end{eqnarray}

\begin{figure}
\begin{center}
  \epsfig{file=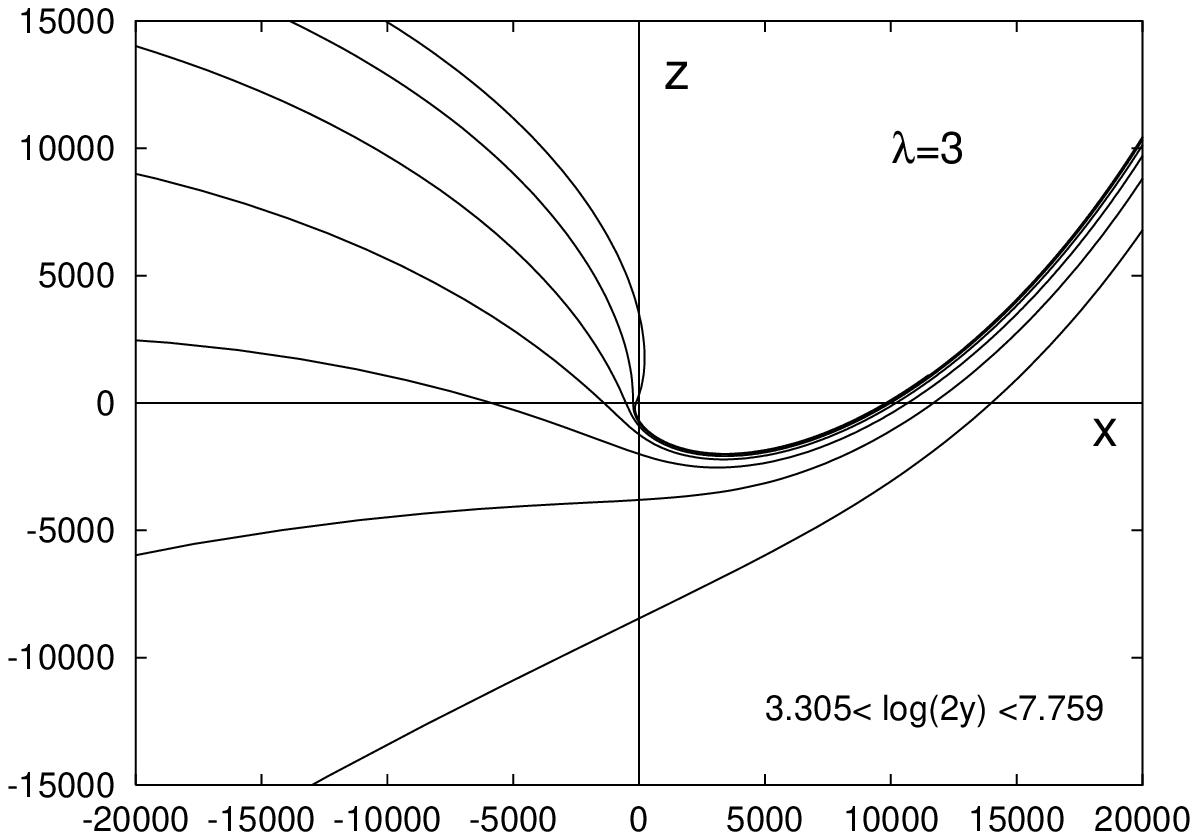,width=85mm} 
  \epsfig{file=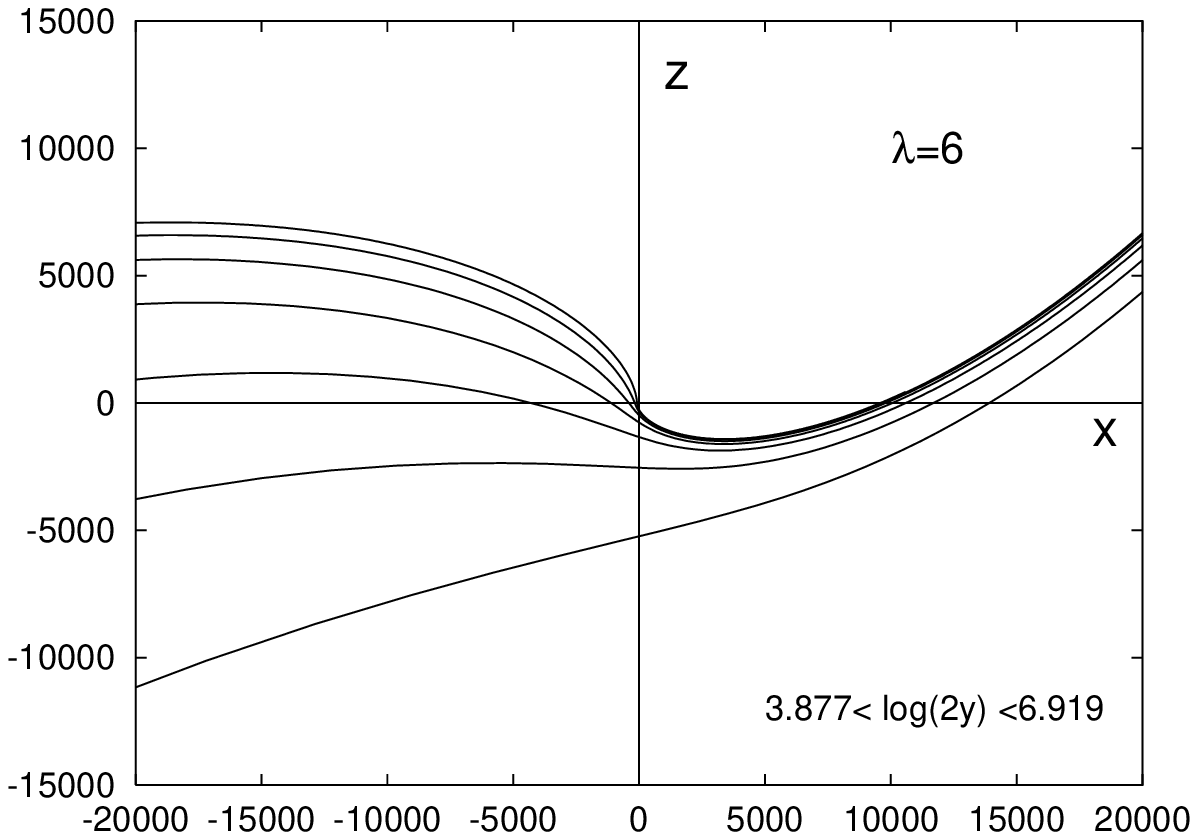,width=85mm}    
\end{center}
\caption{\small Two-branch spirals. The filament
projection is presented for several time moments, $|t_{n+1}/t_n|=1/2$.} 
\label{spirals}
\end{figure}

\end{widetext}

The variable $\rho$ runs here between two neighbour zeros of the function 
$T_{\varkappa^2}(\rho)$ and approaches these values at two logarithmic 
branches of the spiral, $\rho_j^{(\varkappa)}<\rho<\rho_{j+1}^{(\varkappa)}$.

\subsection{The case $F(\rho)=\log\rho$}

For further investigation let us substitute $F(\rho)=\log\rho$ into the
equations (\ref{t_lambda_U}-\ref{U_lambda}) and change the variable 
$$
q=\log\rho.
$$ 
As the result, we will obtain 
\begin{eqnarray}\label{t_lambda_Uq}
t_\lambda&=&\mbox{Re}\{e^q\,U_\lambda'(q)\Theta_\lambda(\vartheta)\}, \\
x_\lambda&=&\mbox{Re}\{
U_\lambda(q)\Theta_\lambda'(\vartheta)\cos\vartheta \nonumber \\
\label{x_lambda_Uq}
&&\qquad+ \left[U_\lambda(q)+U_\lambda'(q)\right]
\Theta_\lambda(\vartheta)\sin\vartheta \},\\
Z_\lambda&=&\mbox{Re}\{
U_\lambda(q)\Theta_\lambda'(\vartheta)\sin\vartheta \nonumber \\
\label{Z_lambda_Uq}
&&\qquad-\left[U_\lambda(q)+U_\lambda'(q)\right]
\Theta_\lambda(\vartheta)\cos\vartheta \} .
\end{eqnarray}
\begin{equation}\label{Uq}
qU_\lambda''(q)+(1+q)U_\lambda'(q)+(1+\lambda)U_\lambda(q)=0.
\end{equation}
General solution of Eq.(\ref{Uq}) is representable by the Laplace method
\cite{LL3} as arbitrary linear combination 
$A_\lambda I_\lambda^{\cal A}(q)+B_\lambda I_\lambda^{{\cal B}}(q)$ 
of two contour integrals in a complex plane,
\begin{eqnarray}
U_\lambda(q)&=&\frac{A_\lambda}{2\pi i}
\oint_{\cal A}\left(\frac{p}{p+1}\right)^{\lambda+1}
\frac{e^{pq}dp}{p} \nonumber \\
\label{Uq_solution}
&&+B_{\lambda}\int_{\cal B}
\left(\frac{p}{p+1}\right)^{\lambda+1}
\frac{e^{pq}dp}{p}.
\end{eqnarray}
Here the first closed contour ${\cal A}$ goes around the points
$p_0=0$ and $p_1=-1$. 
The second contour ${\cal B}$ is not closed, at positive  $q$ it starts
at $\mbox{Re\,}p=-\infty$. If $\mbox{Re\,}\lambda<0$, then ${\cal B}$ ends at
$p_1$, but if $\mbox{Re\,}\lambda\ge 0$, then its end point is $p_0$. 
In both cases at the end point of the contour ${\cal B}$ the integrand
multiplied by $p(p+1)$ tends to zero.

It is interesting to note that at the integer values of the parameter
$\lambda$ the integral $I_\lambda^{\cal A}(q)$  can be expressed in terms 
of polynomials:
\begin{eqnarray}
&&I_\lambda^{\cal A}(q)=\frac{1}{2\pi i}
\oint_{{\cal A}}\left(\frac{p}{p+1}\right)^{\lambda+1}
\frac{e^{pq}dp}{p} \nonumber \\
\label{U_integer_nu}
&&=\left\{
\begin{array}{r}
\left(1+\frac{d}{dq}\right)^{|\lambda|-1}\frac{q^{|\lambda|-1}}{(|\lambda|-1)!},
\quad\lambda=-1,-2,..;\\
e^{-q}\left(\frac{d}{dq}-1\right)^{\lambda}\frac{q^{\lambda}}{\lambda!},
\qquad \quad \lambda = 0,1,2,...
\end{array}\right.
\end{eqnarray}
These expressions have been used to prepare 
Figs.(\ref{k_sqrt9})-(\ref{spirals}) where the vortex filament shape
corresponding to $U_\lambda(q) = I_\lambda^{\cal A}(q)$ is drawn for 
several moments of time. It is easily to see, at sufficiently large times 
the spirals satisfy the conditions a), b), c) that have been formulated 
in the Introduction. As to the angle-shaped configurations, 
the condition $Y'^2\ll X'^2+Z'^2$, generally speaking, 
is not satisfied at $q \gtrsim k^2$, since at very large $q$
(on the asymptotes of the angle) the growth of $Y\sim\exp(q)$ is faster 
than growth of $X,Z\sim U_{-k^2}(q)\sim q^{k^2-1}$.
Therefore, if we take a particular solution 
$u=U_{-k^2}(q)\cos(k\vartheta)$ separately, not as a term in a more complex
linear combination, then we have to deal only with large $k$, and 
consider only the pieces of the filament where  $2..3 \lesssim q \ll k^2$.

\subsection{The case $F(\rho)=\rho^\alpha/\alpha$}

In Ref.\cite{RPR2001PRE} we investigated another regularization of the
Hamiltonian functional that corresponds to $F(\rho)=\rho^\alpha/\alpha$, with
some small positive parameter $\alpha$.
That time we did not see applicability of the hodograph method and therefore
we were able to find only few particular solutions. Now it has been clear that 
in this case a simple substitution exists that reduces
the problem to 2D equation  $\Delta_2f+f=0$. Thus, it becomes possible 
to present a very wide class of solutions of the equation
\begin{equation}\label{equation_for_t_alpha}
\rho^{2}t_{\rho\rho}+2\alpha\rho \,t_{\rho}+
\alpha(1-\alpha)t_{\vartheta\vartheta}=0
\end{equation}
as linear combinations of singular fundamental solutions (that are expressed
through the McDonald function $K_0$) and regular exponential or polynomial
solutions. 
Indeed, by the substitutions
\begin{equation}
\rho=e^q,\quad\vartheta=\phi\sqrt{\alpha(1-\alpha)},
\quad t=e^{(1/2-\alpha)q}f(q,\phi)
\end{equation}
Eq.(\ref{equation_for_t_alpha}) is reduced to the equation with constant
coefficients,
\begin{equation}
f_{qq}+f_{\phi\phi}-(1/2-\alpha)^2f=0.
\end{equation}
As it is well known, the fundamental solutions of this equation have the form
\begin{equation}
f(q,\phi;q_0,\phi_0)=K_0\left(\Big|\frac{1}{2}-\alpha\Big|
\sqrt{(q-q_0)^2+(\phi-\phi_0)^2}\right),
\end{equation}
where $q_0$ and $\phi_0$ are arbitrary parameters. Therefore 
Eq.(\ref{equation_for_t_alpha}) has particular solutions
\begin{equation}
t=\rho^{1/2-\alpha}
 K_0\left(\Big|\frac{1}{2}-\alpha\Big|
\sqrt{\left[\log\frac{\rho}{\rho_0}\right]^2+
\frac{(\vartheta-\vartheta_0)^2}{\alpha(1-\alpha)}}\right).
\end{equation}

It is interesting to note that at $\alpha=1/2$ the system possesses conformal
symmetry. A deep reason of this symmetry is not clear yet.

As concerning separation of the variables,
the function $T_\lambda(\rho)$ in Eqs.(\ref{t_lambda_T})-(\ref{Z_lambda_T})
is given by the expression
\begin{equation}\label{T_alpha}
T_\lambda(\rho)=A_\lambda^+\rho^{s_+(\lambda)}+A_\lambda^-\rho^{s_-(\lambda)}, 
\end{equation}
where $A_\lambda^\pm$ are arbitrary constants. 
The complex exponents $s_\pm(\lambda)$ 
are the roots of the quadratic equation
\begin{equation}
s(s-1)+2\alpha s+\alpha(1-\alpha)\lambda=0.
\end{equation}
Thus,
\begin{equation}
s_{\pm}(\lambda)={1}/{2}-\alpha\pm
\sqrt{\left({1}/{2}-\alpha\right)^2-\alpha(1-\alpha)\lambda}.
\end{equation}

It should be mentioned the solutions presented in \cite{RPR2001PRE} 
correspond to the particular case $\alpha+s=2$.

{~}

\section{Conclusions}
 
In this article an approximate exactly solvable nonlinear model has been 
derived to describe unstable locally-quasi-2D ideal flows with a thin 
vortex filament near a flat boundary. The hodograph method has been applied 
and some particular solutions have been analytically found by variables 
separation in the governing linear partial differential equation for 
auxiliary function $u$. 
More general solutions $u(\rho,\vartheta)$ can be obtained as linear 
combinations of the terms (\ref{u_lambda}) with different 
$\lambda$, but only in few cases it will be possible to resolve analytically 
the dependence $t=-u_\rho(\rho,\vartheta)/F''(\rho)$. 
However, this procedure can be performed numerically.

Though we derived the exactly solvable model under several restrictive 
simplifications, the solutions obtained in this work promise benefit 
in many aspects. For instance, they
may serve as basic approximations in future more advanced analytical 
studies that will take into account effects of non-locality and/or 
finite variable cross-section of the filament, as well as surface waves
in the case of free boundary.

\subsection*{Acknowledgments}
These investigations were supported by INTAS (grant No. 00-00292),
by RFBR , 
by the Russian State Program of Support of the Leading Scientific Schools, 
and by the Science Support Foundation, Russia.

\end{document}